\documentclass[preprint, showpacs,preprintnumbers,amsmath,amssymb,showkeys]{revtex4}

\usepackage{graphicx}
\usepackage{dcolumn}
\usepackage{bm}

\begin{document}

\preprint{APS/123-QED}

\title{Dispersion properties of non-radiating configurations:\\FDTD modeling}

\author{A. D. Boardman}
\author{K. Marinov}%
 \email{k.marinov@salford.ac.uk}
\affiliation{%
Photonics and Nonlinear Science Group, Joule Laboratory, Department of Physics,
University of Salford, Salford M5 4WT, UK
}%

\author{N. Zheludev}
 \homepage{http://www.nanophotonics.phys.soton.ac.uk}
\author{V. A. Fedotov}
\affiliation{%
EPSRC Nanophotonics Portfolio Centre, School of Physics and
Astronomy, University of Southampton, Highfield, Southampton, SO17
1BJ, UK
}%

\date{\today}

\begin{abstract}
A finite-difference time-domain (FDTD) numerical analysis is used to demonstrate that a
toroidal solenoid, coaxial with an electric dipole, is a remarkable
non-radiating configuration. It can be used to measure the
dielectric permittivity of any ambient matter. It becomes a directional
radiator at an interface between two dielectric media, depositing energy
in the material with the highest polarizability.
\end{abstract}

\maketitle

\section{\label{sec:level1}Introduction}

Toroidal and supertoroidal structures are widely present in nature
and a supertoroid was explicitly drawn by Leonardo in 1490. The
simplest examples of such objects would be toroidal solenoids with
currents in them. More generally, fractal complications of the
simple toroidal wiring known as supertoroidal structures are
discussed and toroidal arrangements of electric and magnetic
dipoles have been discussed in the literature. Today the main
biological journals feature an increasing number of papers on
proteins, viruses and phages possessing elements of toroidal and
supertoroidal symmetry. At the same time we witness a growing
stream of theoretical papers on the electrodynamics and optics of
toroidal and supertoroidal currents, toroidal nanostructures,
toroidal microscopic moments and interactions of electromagnetic
fields with toroidal configuration \cite{BoardmanAndZheludev04}. Recent studies
of phase transitions in ferroelectric nanodisks and nanorods
\cite{NaumovBellaicheAndFu04} and toroidal arrangements of spins
in magnetic structures \cite{KlauiEtAl03} show growing
interest in studying toroidal structures from the materials
research community.

Here we report for the first time a rigorous finite-difference time-domain numerical analysis proving that a toroidal solenoid with
poloidal wiring coaxial with an electric dipole is a remarkable
non-radiating configuration. The property not to radiate
electromagnetic energy is based on the destructive interference
between the fields created by each of its constituents.  We show
that this configuration may be used as a sensor for the dielectric
permittivity of the ambient matter. It becomes a directional
radiator at an interface between two dielectric media depositing
energy in the material with highest polarizability.

Non-radiating configurations are such oscillating charge-current
distributions that do not produce electromagnetic fields in the
radiation zone. An early work \cite{schott33} shows that the orbital
motion of a uniformly charged spherical shell of radius $R$ will
not produce any radiation if the radius $R$ of the shell is equal
to $lcT/2$ where $c$ is the speed of light, $T$ is the period of
the orbit and $l$ is an integer number.
The general problem for absence of radiation from an arbitrary
localized charge distribution, exhibiting periodic motion with
period $T=2\pi/\omega_{0}$, has been addressed \cite{goedecke64} and 
it has been shown that such a system does not generate electromagnetic potentials
in the radiation zone if the Fourier components
$\widetilde{J}(l\omega_{0}\bm{r}/cr,l\omega_{0})$ are not present in the
spectrum of the current density $\bm{J}(\bm{r},t)$. This
criterion also explains the behavior of an orbiting uniformly charged sphere. It
has been pointed out in \cite{goedecke64} that this condition may
not be necessary. It indeed ensures the disappearance of the
electromagnetic potentials in the radiation zone, however calculations 
of the power emitted by the system show that its value is
zero provided that
$\widetilde{J}(l\omega_{0}\bm{r}/cr,l\omega_{0})\propto\bm{r}$, which is a
weaker sufficient condition. Indeed, the latter condition only
requires the absence of the
components transverse to the wave-vector. It
has been proved rigorously \cite{DevaneyWolf73} that the
absence of the transverse components of the Fourier spectrum of
the current density is a necessary and sufficient condition for
absence of radiation. Interestingly, such a condition has appeared
in an earlier study, \cite{BohmAndWeinstein48}, in connection with
electromagnetic self-force action and self-oscillations of a
non-relativistic particle.

The important conclusion that can be drawn from the earlier results 
is that two types of non-radiating configurations can exist in principle. For the
first type the Fourier components
$\widetilde{J}(\omega\bm{r}/cr,\omega)$ of the current density
are zero. Numerous examples of systems pertaining to this sort of
non-radiating configurations exist - \cite{schott33, goedecke64,
AfanasievAndDubovik98}. A characteristic feature of these systems
is that \textit{both} the electromagnetic fields and the
electromagnetic potentials are zero. For the second type of
non-radiating configurations the Fourier spectrum is purely longitudinal
i.e. $\widetilde{J}(\omega\bm{r}/cr,\omega)\propto\bm{r}$.
Here the electromagnetic fields are zero but as we show the 
electromagnetic potentials may
be finite. 

Interestingly, it is pointed out \cite{goedecke64} that the case of
$\widetilde{J}(\omega\bm{r}/cr,\omega)\propto\bm{r}$ corresponds
to trivial spherically symmetric radial oscillations of the charge
density. Nevertheless non-trivial examples can be created using toroidal structures. 
Recent papers \cite{AfanasievAndDubovik98}
and \cite{AfanasievAndStepanovsky95} show that a
non-radiating configuration can be constructed by combining an
infinitesimal toroidal solenoid with poloidal current flowing in
its windings (i.e. along the meridians of the toroid) with an
electric dipole placed in the center of the toroid. The explicit
calculations of \cite{AfanasievAndDubovik98} and
\cite{AfanasievAndStepanovsky95} show that while the
electromagnetic fields disappear outside such a composite object,
the electromagnetic potentials survive. As we show here this particular
structure belongs to the second type of non-radiating systems and
that it is the longitudinal part of the Fourier-spectrum of the current density 
which is responsible for the residual electromagnetic potentials in the radiation zone.

The results of \cite{AfanasievAndDubovik98} and
\cite{AfanasievAndStepanovsky95} suggest that the non-radiating
configurations involving toroidal solenoids may have a number of interesting electromagnetic properties. These properties however have never been
studied in proper detail. This is the aim of the present study. 

The physical nature of the problem is extremely well-suited  to
numerical modelling using the FDTD method
\cite{TafloveAndHagness2000} which will be used in our analysis.
It gives the possibility to address the electromagnetic
properties of this specific structure consisting of a toroid
coupled to a dipole in full numerical detail. In addition, an
assessment of what possible applications such structures might
have is given.

Exact compensation  between the fields generated by a toroidal
solenoid and an electrical dipole takes place for infinitesimal
objects only. It therefore seems plausible that assessment of the
extent to which the properties of the infinitesimal non-radiating
configurations are preserved by finite-dimensional
counterparts should precede possible experimental designs. It is
important also to determine what is the behavior of these
structures under certain (non-trivial) perturbations.

\section{Infinitesimal toroidal solenoids and non-radiating configurations}

The electromagnetic properties of toroidal solenoids and toroidal helix antennas are studied in detail in references \cite{Afanasiev93,
Afanasiev90, AfanasievAndDubovik92, AfanasievDubovikMisicu93, Afanasiev01, DubovikAndTugushev90, LeWeiLi04}. Here only the results that will be
used in our exposition are briefly summarized.
\begin{figure}
\includegraphics{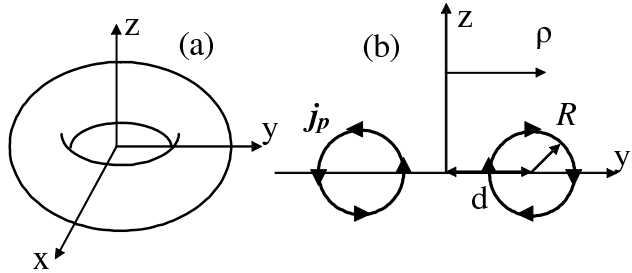} 
\caption{\label{fig:fig1.} Toroidal surface (a) and its cross-section with the $z-y$ plane (b). The triangles show the direction of the
surface current $\bm{j}_{p}$ .}
\end{figure}

The current flowing along the meridians of a toroidal solenoid (known also as poloidal current, Fig. 1) can be presented in the form
(see e.g. \cite{Afanasiev01}):
\begin{eqnarray}
    \bm{j_{p}}=\nabla\times\bm{M},
    \label{eq:one}
\end{eqnarray}
since $\nabla.\bm{j_{p}}=0.$ In Eq.~(\ref{eq:one}) $\bm{j_{p}}$ is the current density vector and $\bm{M}=(0, M_{\varphi}, 0)$  is
the azimuthal magnetization vector, given by
\begin{eqnarray}
        \bm{M_{\varphi}}=\frac{NI(t)}{2\pi\rho}
    \label{eq:two}
\end{eqnarray}
if $(\rho-d)^{2}+z^{2}<R^{2}$ and zero otherwise. In Eq.~(\ref{eq:two}) $N$ is the total number of windings and $I[A]$ is the magnitude
of the current. Pursuing this idea a step further the magnetization $\bm{M}$ can in turn be expressed as
\begin{eqnarray}
    \bm{M}=\nabla\times\bm{T},
    \label{eq:three}
\end{eqnarray}
where $\bm{T}=(0, 0, T_{z})$ is called toroidization vector. The general problem is difficult to perform analytically
\cite{Afanasiev93, Afanasiev90, AfanasievDubovikMisicu93, AfanasievAndDubovik92} and any limit that preserves the correct properties
while at the same time giving valuable mathematical simplification is a step worth taking. Such a step is  $d\rightarrow0$. This is
a useful step because it gives the toroidization vector in the following form (see e.g. \cite{AfanasievAndDubovik98, Afanasiev01}).
\begin{eqnarray}
    T_{z}=\frac{\pi N I d R^{2}}{2}\delta^{3}(r).
    \label{eq:four}
\end{eqnarray}
Assuming monochromatic time-dependence, $\propto\exp(-i\omega t)$, and using (4) the magnetic field created by the toroidal solenoid
can be obtained in the form
\begin{eqnarray}
    \bm{H}_{p}=\frac{ N I d R^{2}}{8}\frac{k^{2}}{r^{2}}(ik-\frac{1}{r})(\bm{r}\times\bm{n})\exp(ikr),
    \label{eq:five}
\end{eqnarray}
where $\bm{n}$ is a vector of unit length pointing along the $z$-axis and $k$ is the wave vector.

A dipole can be introduced at the center of the toroid. If this dipole is modeled as a piece of wire of length $L_{d}$  with the
current strength being equal to $I_{d}$ the dipole moment amplitude, $\bm{p}_{0}$, can be expressed through
$iL_{d}I_{d}=\omega p_{0}$, where  $\bm{p}_{0}=p_{0}\bm{n}$. The magnetic field of the dipole is \cite{Jackson99}
\begin{eqnarray}
    \bm{H}_{d}=\frac{\omega k}{4\pi}(1-\frac{1}{ikr})(\bm{r}\times\bm{p}_{0})\frac{\exp(ikr)}{r^{2}}
    \label{eq:six}
\end{eqnarray}
The time-averaged power $P$ emitted by the composite object (an infinitesimal toroidal solenoid coupled to an electrical dipole) is given by
\begin{eqnarray}
    P=\frac{\mu_{0}ck^{2}}{12\pi\sqrt{\epsilon}}(I_{d}L_{d}+k^{2}T)^{2},
    \label{eq:seven}
\end{eqnarray}
where $T=\pi NIdR^{2}/2$ and $\epsilon$ is the relative dielectric permittivity of the ambient matter.

This expression can be generalized to include  higher-order multipole moments \cite{DubovikAndTugushev90}.

Equation~(\ref{eq:seven}) can be rewritten in the form
\begin{eqnarray}
P=\frac{\mu_{0}ck^{2}(I_{d}L_{d})^{2}}{12\pi\sqrt{\epsilon}}\left(1-\frac{\epsilon}{\widetilde{\epsilon}}\right)^{2}
    \label{eq:eight}
\end{eqnarray}
where
\begin{eqnarray}
\widetilde{\epsilon}=-\frac{I_{d}L_{d}c^{2}}{\omega^{2} T}
\label{eq:nine}
\end{eqnarray}
is the effective relative dielectric permittivity of the medium in which electromagnetic fields of the toroid and the electric dipole
can compensate each other. This suggests that it should be possible to measure the relative dielectric permittivities of media
(e.g. liquids) by adjusting experimentally the ratio of the currents $I_{d}$ and $I$  until a minimum of the emitted power is detected.
Then the relative dielectric constant of the material under investigation can be obtained from (\ref{eq:nine}).

It has been pointed out in \cite{AfanasievAndStepanovsky95} that while the electromagnetic fields disappear when the compensation
condition (\ref{eq:nine}) is satisfied the electromagnetic potentials survive. However, there are examples of non-radiating
configurations (see e.g. \cite{goedecke64, AfanasievAndDubovik98}) for which \textit{both} the electromagnetic fields and the
electromagnetic potentials are zero. The question is then what is the physical reason for that and what is the difference between
both types of electromagnetic systems. Following \cite{goedecke64, DevaneyWolf73} it can be shown that the difference is in the
current-density spectra. To see this consider the vector potential
\begin{eqnarray}
\bm{A}=\frac{\mu_{0}}{4\pi}\int\frac{\bm{j}(\bm{r'},t-\frac{\left|\bm{r}-\bm{r'}\right|}{c})}{\left|\bm{r}-\bm{r'}\right|}d^{3}\bm{r'}.
 \label{eq:ten}
\end{eqnarray}
In the radiation zone the standard approximation  \cite{Jackson99}  can
be used and (\ref{eq:ten}) reduces to
\begin{eqnarray}
    \bm{A}=\frac{\mu_{0}}{4\pi r}\int\bm{j}(\bm{r'},t-r/c+\bm{r}.\bm{r'}/cr)d^{3}\bm{r'}
    \label{eq:eleven}
\end{eqnarray}
Now if the current density $\bm{j}(\bm{r},t)$ is expressed through its Fourier-transform
\begin{eqnarray}
    \bm{j}(\bm{r},t)=\int\bm{\widetilde{j}}(\bm{k},\omega)e^{-i(\omega t-\bm{k}.\bm{r})}d^{3}\bm{k}d\omega
    \label{eq:twelve}
\end{eqnarray}
Eq. (\ref{eq:eleven}) becomes
\begin{eqnarray}
    \bm{A}=\frac{\mu_{0}(2\pi)^{3}}{4\pi r}\int\bm{\widetilde{j}}(\frac{\omega\bm{r_{0}}}{c},\omega)e^{-i\omega( t-r/c)}d\omega
 \label{eq:thirteen}
\end{eqnarray}
where $\bm{r}_{0}=\bm{r}/r$. As Eq. (\ref{eq:thirteen}) shows only those components of the current density spectrum that correspond
to $|\bm{k}|=\omega/c$ contribute to radiation \cite{DevaneyWolf73}.  It is immediately clear that if the condition
\begin{eqnarray}
    \bm{\widetilde{j}}(\frac{\omega\bm{r_{0}}}{c},\omega)=0
    \label{eq:fourteen}
\end{eqnarray}
is satisfied then the vector potential vanishes. Using the continuity equation, the Fourier-components of the charge density can be
expressed from the Fourier-components of the current density according to  $\rho(\bm{k},\omega)=\bm{k}.\bm{\widetilde{j}}(\bm{k},\omega)/\omega$
and by following a procedure similar to deriving Eq. (\ref{eq:thirteen}) but this time for the scalar potential it can be shown that the
scalar potential is also zero if (\ref{eq:fourteen}) is satisfied. Therefore (\ref{eq:fourteen}) ensures that the electromagnetic system
considered is a non-radiating configuration. This general statement is a sufficient condition \cite{goedecke64}.

The results of \cite{DevaneyWolf73} imply however that the condition (\ref{eq:fourteen}) is \textit{not} necessary. With the assumption of a monochromatic
time-dependence (\ref{eq:thirteen}) reduces to
\begin{eqnarray}
    \bm{A}=\frac{\mu_{0}(2\pi)^{3}}{4\pi r}\bm{\widetilde{j}}(\frac{\omega\bm{r_{0}}}{c},\omega)e^{-i\omega( t-r/c)}.
 \label{eq:fiveteen}
\end{eqnarray}
The electromagnetic fields can then be obtained using $\bm{H}=\nabla\times\bm{A}/\mu_{0}$ and $\bm{E}=i\nabla\times\bm{H}/\omega\epsilon_{0}$.
Beacause (\ref{eq:fiveteen}) is valid in the radiation zone only, $\bm{r}_{0}$ and $1/r$ must be treated as constants in deriving the fields from the vector potential. The result is
\begin{eqnarray}
    \bm{E}=i\sqrt{\frac{\mu_{0}}{\epsilon_{0}}}\frac{\omega (2\pi)^{3} }{4\pi c r}\bm{r}_{0}\times(\bm{\widetilde{j}}\times\bm{r}_{0})e^{-i\omega(t-r/c)}
 \label{eq:sixteen}
\end{eqnarray}
and
\begin{eqnarray}
\bm{H}=i\frac{\omega (2\pi)^{3}}{4\pi c r}(\bm{\bm{r}_{0}\times\widetilde{j}})e^{-i\omega(t-r/c)}.
\label{eq:seventeen}
\end{eqnarray}
From (\ref{eq:sixteen}) and (\ref{eq:seventeen}) it is clear that the
time-averaged Poynting vector, $\left\langle \bm{S} \right\rangle=\frac{1}{2}\bm{E}\times\bm{H}^{*}$, can be presented in the form
\begin{eqnarray}
    \left\langle \bm{S} \right\rangle\propto|\bm{r}_{0}\times(\bm{\widetilde{j}}\times \bm{r}_{0})|^{2}\bm{r}_{0}
    \label{eq:eighteen}
\end{eqnarray}
The quantity $\bm{r}_{0}\times(\bm{\widetilde{j}}\times \bm{r}_{0})$ is the radiation pattern of the system. As can be seen from (\ref{eq:eighteen}) a charge-current distribution will not emit electromagnetic energy if
\begin{eqnarray}
    \bm{\widetilde{j}}_{\bot}\equiv\bm{r}_{0}\times(\bm{\widetilde{j}}\times \bm{r}_{0})=0
    \label{eq:nineteen}
\end{eqnarray}
which is a weaker sufficient condition compared to (\ref{eq:fourteen}). The fact that (\ref{eq:nineteen}) is also a necessary condition
for the absence of radiation can be seen by setting $\bm{E}$ and $\bm{H}$ to zero in (\ref{eq:sixteen}) and (\ref{eq:seventeen}) and this was established in \cite{DevaneyWolf73}. However (\ref{eq:nineteen}) has appeared in the earlier studies \cite {BohmAndWeinstein48, goedecke64}.

The identity $\bm{\widetilde{j}}=\bm{r}_{0}\times(\bm{\widetilde{j}}\times \bm{r}_{0})+\bm{r}_{0}(\bm{r}_{0}.\bm{\widetilde{j}})=\bm{\widetilde{j}}_{\bot}+\bm{\widetilde{j}}_{||}$ and the comparison of (\ref{eq:nineteen}) with (\ref{eq:fourteen}) show that systems satisfying (\ref{eq:fourteen}) will emit no
electromagnetic energy and will not produce electromagnetic potentials. On the other hand for systems satisfying the weaker
condition (\ref{eq:nineteen}) the electromagnetic potentials are not necessarily zero, since the longitudinal (parallel to $\bm{r}_0$)
part of the current-density spectrum $\bm{\widetilde{j}}_{||}$ will contribute to the vector potential according to Eq. (\ref{eq:fiveteen}). 

It is easy to show that the non-radiating configuration consisting of a toroidal solenoid coaxial with an electric dipole is an
example of the second type of non-radiating configurations - systems satisfying (\ref{eq:nineteen}) and not (\ref{eq:fourteen}).
The current density associated with this system is $\bm{j}(\bm{r})=T\nabla\times(\nabla\times\bm{n}\delta^{3}(\bm{r}))+I_{d}L_{d}\bm{n}\delta^{3}(\bm{r})$ and its Fourier spectrum is given by
\begin{eqnarray}
    (2\pi)^{3}\bm{\widetilde{j}}(\bm{k})=-T\bm{k}(\bm{k}.\bm{n})+(k^{2}T+I_{d}L_{d})\bm{n}
    \label{eq:twenty}
\end{eqnarray}
For values of $\bm{k}=\omega\sqrt{\epsilon}\bm{r}_{0}/c$ and using the compensation condition (\ref{eq:nine}), Eq. (\ref{eq:twenty}) reduces to
\begin{eqnarray}
(2\pi)^{3}\bm{\widetilde{j}}(\frac{\omega\bm{r_{0}}}{c})=-T\frac{\omega^{2}\epsilon}{c^{2}}\bm{r}_{0}(\bm{r_0}.\bm{n}).
    \label{eq:twenty_one}
\end{eqnarray}
As (\ref{eq:twenty_one}) shows the current density spectrum is purely parallel to $\bm{r}_{0}$ for wavenumber values $|\bm{k}|=\omega\sqrt{\epsilon}/c$.
It can be concluded that it is the survival of the longitudinal part of the current density spectrum that gives the possibility to create
non-zero electromagnetic potentials in the radiation zone in the absence of electromagnetic fields.

\section{Numerical modeling of the interaction of a non-radiating configuration with the interface between two materials}
\begin{figure}
\includegraphics{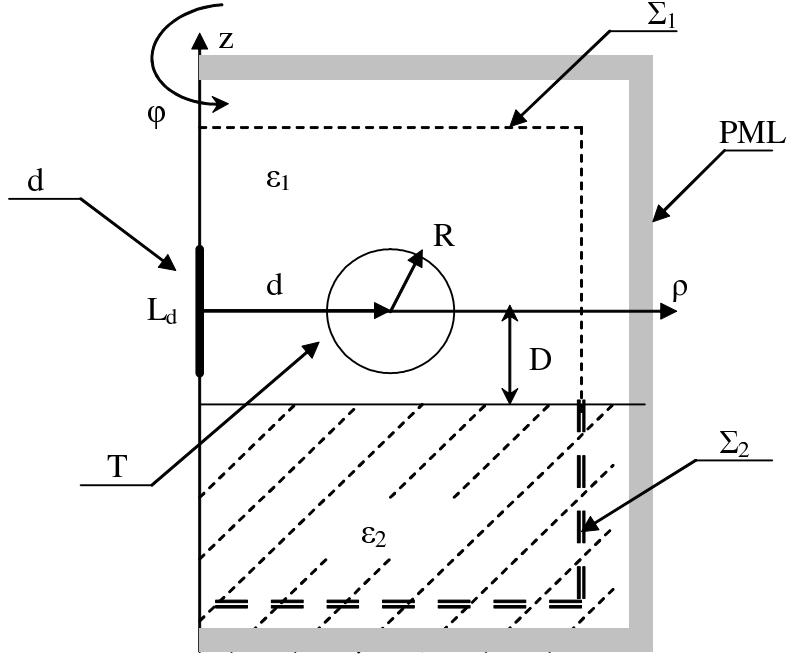}
\caption{\label{fig:fig2} Non-radiating configuration consisting of a toroidal solenoid and an electric dipole near the interface
between two materials. Cylindrical symmetry is assumed. PML - perfectly matched layer; $d$ - dipole, $T$- toroid; $\Sigma_{1}$, $\Sigma_{2}$ -
cylindrical surfaces used to calculate the power emitted in each of the materials with $\epsilon=\epsilon_{1}$  and  $\epsilon=\epsilon_{2}$,
respectively.}
\end{figure}
As can be seen from Section 2 the composite emitter - toroid and dipole - becomes a non-radiating configuration (note that the
compensation condition (\ref{eq:nine}) is satisfied) due to the destructive interference between the electromagnetic fields created
by the toroid and by the electric dipole. This interference occurs in all possible directions in a homogeneous medium. In
an \textit{inhomogeneous} material however as would be encountered for an interface between two dielectrics with relative
permittivity constants $\epsilon_{1}$, and $\epsilon_{2}$   this does not happen. To appreciate this consider the situation
shown in Figure \ref{fig:fig2}, in which an emitter consisting of an electric dipole and a toroidal solenoid is placed in a
medium with dielectric permittivity equal to $\epsilon_{1}$. This medium is separated from a second one by a planar interface
located at a distance $D$ from the equatorial plane of the toroid. In the absence of the interface the system is non-radiative
and the effective permittivity is $\widetilde{\epsilon}=\epsilon_{1}$. In order to assess the consequences flowing from the
presence of the interface between the two dielectrics and also the finite size of both the toroid and the dipole it is necessary
to solve the Maxwell's equation exactly. This can be achieved numerically using the FDTD method \cite{TafloveAndHagness2000}.
The latter can be considerably simplified since the toroid is a body of revolution (BOR). Taking advantage of the axial symmetry
reduces the problem to a two-dimensional one. Cylindrical coordinates can be used and there is no dependence on the azimuthal
variable angle $\varphi$ (Fig. \ref{fig:fig2}). This implementation of the FDTD method is known as BOR-FDTD \cite{TafloveAndHagness2000}.
The computational domain is terminated by a standard perfectly matched layer (PML) \cite{TafloveAndHagness2000, TeixeiraAndChew97}.
The radiation of both the toroid and the dipole is categorized by the field components $(E_{\rho}, E_{z}, H_{\varphi})$  that are
not identically zero and hence it is of E-type (TM) \cite{Afanasiev01}. The applicability of the FDTD method to radiating
structures (antenna problems) is well established and this technique has been successfully applied to various designs
\cite{MaloneySmithAndScott90, MaloneyAndSmith93, TircasAndBalanis92}. In the model the poloidal current $\bm{j_{p}}$,
is expressed through the azimuthal component of the magnetization which is consistent with the assumption that all the
parts of the toroid respond simultaneously (or with negligible delay) to the driving voltage. This is expected to occur
when the size of the toroid is much smaller than the wavelength.
To evaluate the directional properties of the emitting structure studied, the quantities $P_{1}$ and $P_{2}$ are introduced
and defined as
\begin{eqnarray}
    P_{i}=\int_{\Sigma_{i}}\left\langle\bm{S}\right\rangle.d\bm{\Sigma_{i}}, i=1, 2
    \label{eq:twenty_two}
\end{eqnarray}
In (\ref{eq:twenty_two}) $\left\langle\bm{S}\right\rangle$ is the time-averaged Poynting vector and $\Sigma_{1}$, $\Sigma_{2}$
(see Fig. \ref{fig:fig2}) are cylindrical surfaces placed away from the source (close to the PML region) in order to ensure
that the near-field contributions have negligible effect on the power values calculated according to (\ref{eq:twenty_two}).

\section{Results and discussion}

\begin{figure}
\includegraphics{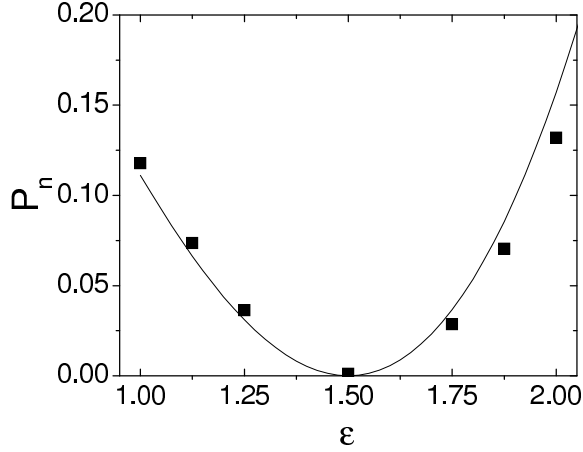}
\caption{\label{fig:fig3} Normalized emitted power $P_{n}=12P\pi c/(\mu_{0}(I_{d}L_{d}\omega)^{2})$  versus the relative dielectric
permittivity $\epsilon$ of the ambient dielectric material. The value of $\widetilde{\epsilon}$  (Eq. (\ref{eq:nine}))
is  $\widetilde{\epsilon}=1.5$. The solid curve and the solid squares are the analytical result (Eq. (\ref{eq:eight})) and
the numerical result, respectively.}
\end{figure}

In order to model the behavior of the non-radiating configuration the following parameter values are selected. The larger and the
smaller radii of the toroidal solenoid are fixed to $d=1$ cm and $R=0.5$ cm, respectively, the dipole length is $L_{d}$=0.9 cm
and the excitation frequency is $\omega/2\pi=1$ GHz. The FDTD-grid resolution is $\Delta\rho=\Delta z=\lambda/300$, where $\lambda$
is the free-space wavelength. Since Eq. (\ref{eq:eight}) is strictly valid for infinitesimal objects only, it is necessary to make
sure that for the selected values of the parameters the contributions from the higher-order multipoles are negligible. To verify
this Eq. (\ref{eq:eight}) has been compared with results obtained from FDTD simulations in a homogeneous material (this pertains
to the case of $\epsilon_{1}=\epsilon_{2}=\epsilon$ in Fig. \ref{fig:fig2}) and the result is presented in Figure \ref{fig:fig3}.
The simulations are in good agreement with Eq. (\ref{eq:eight}). This means that for the selected values of the parameters the
contributions of the toroidal dipole moment and the electrical dipole moment are dominant. 

Figure 4 compares the directional
properties of a \textit{perturbed} non-radiating configuration with that of an electric dipole.
\begin{figure}
\includegraphics{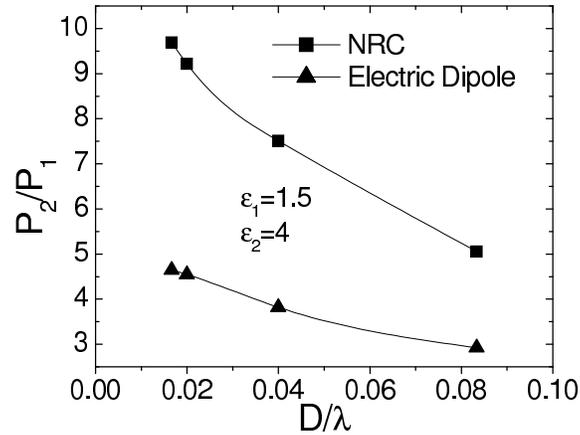}
\caption{\label{fig:fig4} The ratio between the powers $P_{1}$ and $P_{2}$ emitted in the materials with dielectric constant $\epsilon_{1}$
and $\epsilon_{2}$, respectively, by a non-radiating configuration (NRC, solid squares) and an electric dipole (Electric dipole, solid
triangles) as a function of the distance $D$ between the emitter and the interface. The parameter $\widetilde{\epsilon}$ of the
non-radiating configuration is $\widetilde{\epsilon}=\epsilon_{1}$.}
\end{figure} The ratio between the power values $P_{1}$ and $P_{2}$ emitted in the materials with dielectric constants $\epsilon_{1}$
and $\epsilon_{2}$, is computed for several values of the distance $D$ using either a non-radiating configuration or an
electric dipole. 
The presence of the interface affects both types of emitters. However Fig. \ref{fig:fig4} shows that a larger fraction of the total emitted power can be directed in the material with $\epsilon=\epsilon_{2}$ for the case in which
the emitter is a non-radiating configuration. Comparing the performance of the non-radiating configuration with that of the
dipole acting along shows that the non-radiating configurations has a clear advantage in the ability to direct a larger
fraction of the total emitted power in a material with higher value of the dielectric constant. This advantage disappears in
the proportion of the increase of the distance to the interface. It has been verified that the dependence of the ratio $P_{2}/P_{1}$
on the distance $D$ for both the emitters stems mainly from the dependence of the quantity $P_{2}$ on $D$. The value of $P_{1}$
appears to be less susceptible to the variations of $D$ for this range of parameter values.
For some applications it might be desirable to direct electromagnetic energy within a certain material while little or no energy
is emitted to the surrounding space. It seems that a non-radiating configuration with $\widetilde{\epsilon}=\epsilon_{1}$  may
be suitable for this purpose. Relatively far from the interface it does not radiate at all, or radiates a small amount of power.
Bringing the non-radiating configuration into contact with the interface will lead to an increase of the total emitted
power $P_{1}+P_{2}$, (keeping the values of the currents $I_{d}$  and $I$ fixed) with the contribution $P_{2}$  predominating strongly.

\begin{figure}
\includegraphics{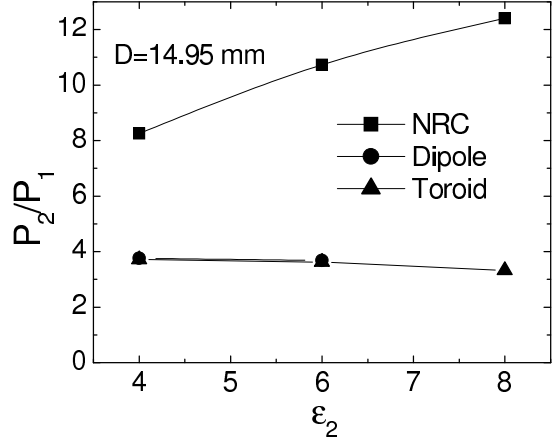}
\caption{\label{fig:fig5} The ratio between the powers $P_{1}$ and $P_{2}$  emitted in the materials with dielectric constant
$\epsilon_{1}$ and $\epsilon_{2}$, respectively, by a non-radiating configuration (NRC, solid squares),  electrical dipole
(Dipole, solid circles) and the toroidal solenoid (Toroid, solid triangles) as a function of the dielectric constant $\epsilon_{2}$. The
parameter $\widetilde{\epsilon}$  of the non-radiating configuration is  $\widetilde{\epsilon}=\epsilon=1$ and the distance between the emitters and the interface is $D=14.95$ mm.}
\end{figure}

\begin{figure}
\includegraphics{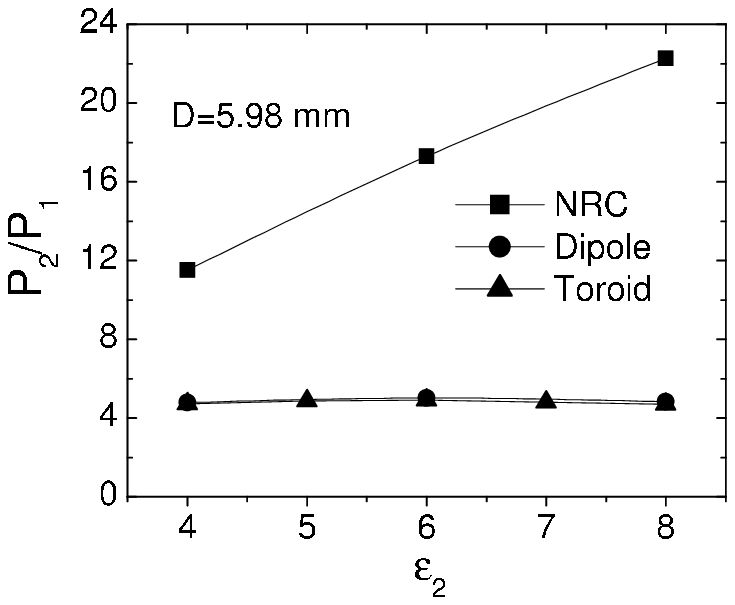}
\caption{\label{fig:fig6} The same as in Figure \ref{fig:fig5} but for $D=5.98$ mm.}
\end{figure}

To study this property further Figure \ref{fig:fig5} and Figure \ref{fig:fig6} show the dependence of the ratio $P_{2}/P_{1}$  on the dielectric constant
of the substrate for two fixed values of the distance $D$ between the emitters and the interface and $\widetilde{\epsilon}=\epsilon_{1}$.
The directional properties of the non-radiating composite object are compared with those of its constituents - the electrical dipole
and the toroidal solenoid. As Fig. \ref{fig:fig5} and Fig. \ref{fig:fig6} show, the ratio  $P_{2}/P_{1}$ for the toroidal solenoid and for the electrical
dipole shows little dependence on the dielectric constant of the substrate $\epsilon_{2}$. At same time, when non-radiating
configuration is used as an emitter, not only the ratio $P_{2}/P_{1}$ is higher, but it increases strongly with the increase
of  $\epsilon_{2}$. This shows that in the region of parameter values studied the directional properties of the non-radiating
configuration improve with the increase of the contrast between the relative dielectric permittivities of the two materials. A
comparison between Fig. \ref{fig:fig5}  and Fig. \ref{fig:fig6} shows that as the  non-radiating configuration approaches the interface
its performance improves. Indeed, it can be concluded that the optimum performance is achieved when the non-radiating configuration
is in direct contact with the interface. This feature is in agreement with Figure \ref{fig:fig4}. The dependence of the emission
properties of the non-radiating configuration upon the values of the dielectric constant of the substrate is suitable for sensor applications.

\begin{figure}
\includegraphics{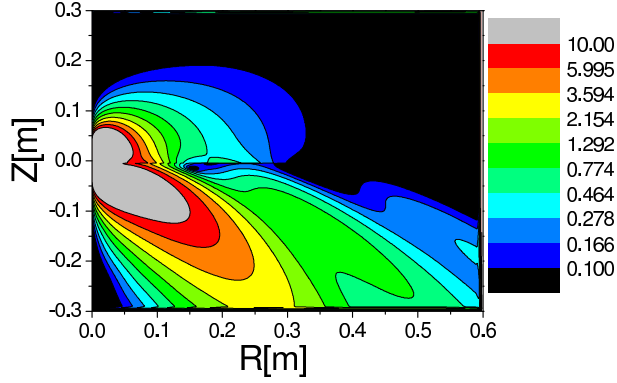}
\caption{\label{fig:fig7}(Color online) Spatial distribution of the time-averaged and normalized Poynting vector modulus
$\left\langle\sqrt{S^{2}_{\rho}+S^{2}_{z}}\right\rangle/(P_{1}+P_{2})$  for a non-radiating configuration. The distance between the emitter and the interface is $D=5.98$ mm. Note that logarithmic
scale is used for the values of the Poynting vector. The values of the other parameters are $\widetilde{\epsilon}=\epsilon_{1}=1$
and $\epsilon_{2}=8$.}
\end{figure}

\begin{figure}
\includegraphics{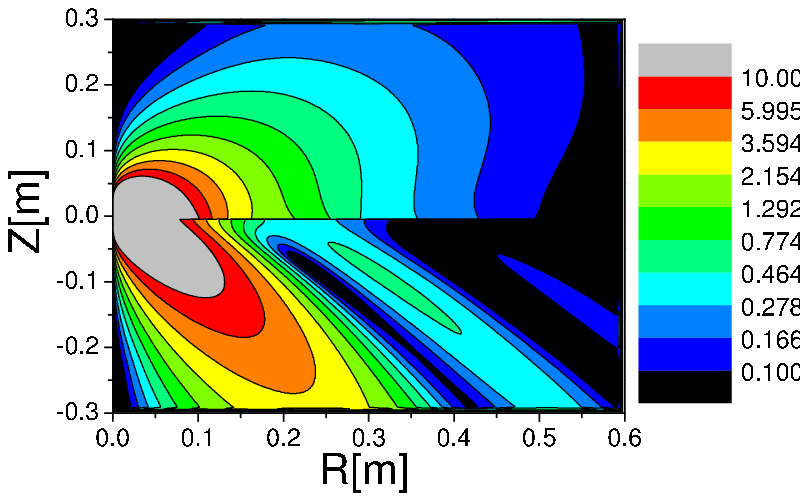}
\caption{\label{fig:fig8}(Color online) The same as in Figure \ref{fig:fig7} but the emitter here is a toroid.}
\end{figure}
The results presented in Fig. \ref{fig:fig6} are  visually
presented in Fig. \ref{fig:fig7} and Fig. \ref{fig:fig8} where the spatial distribution
of the time-averaged Poynting vector around two of the studied
emitters - non-radiating configuration and a toroidal solenoid -
is shown. The Poynting vector values are normalized to the value
of the total emitted power to allow a comparison at identical
total emitted powers to be made. As can be seen the
electromagnetic field created by the non-radiating configuration
in free space ($\epsilon=1$) is weak compared to that emitted by
the toroidal solenoid. This shows that a larger fraction of the
total emitted power is deposited in the substrate material.

\section{Conclusions}

In conclusion we studied a remarkable non-radiating configuration
consisting of a toroidal solenoid coupled to an electrical dipole.
The property not to radiate electromagnetic energy is based on the
destructive interference between the fields created by each of its
constituents. We show that the interference effect depends on the
dielectric characteristics of the ambient matter and the
configuration  may be used in  dielectric
permittivity measurements. It becomes a directional
radiator at an interface between two dielectric media depositing
energy in the material with the highest polarizability.

\section{Acknowledgments}

We acknowledge fruitful discussions on the subject with
J. A. C. Bland, G. Afanasiev,  A. Ceulemans, E. Tkalya, H. Schmid, M.
Martsenyuk and A. Dereux. This work is supported by the
Engineering and Physical Sciences Research Council (UK) under the Adventure
Fund Programme.

\end{document}